# Metal-insulator transition in electric field: A viewpoint from the switching effect

P. P. Boriskov, A. L. Pergament, A. A. Velichko, G. B. Stefanovich and N. A. Kuldin

*Petrozavodsk State University, Petrozavodsk, 185910, Russian Federation*

(Dated: February 28, 2006)

The proposed switching mechanism is based on an electronically-induced metal-insulator transition occurring in conditions of the excess non-equilibrium carrier density under the applied electric field. First, this mechanism is developed on the basis of a phenomenological approach. This model not only allows the qualitative description of the switching mechanism, but it is in quantitative agreement with the experimental results, in particular, with those concerning the critical concentration and threshold field. The mechanism takes into account the dependence of the carrier density on electric field, as well as the scaling of the critical field. Next, we show that such a "macroscopic" approach can be supported by some microscopic model. The quintessence of this approach consists in the fact that an electronically induced metal-insulator transition may be described in terms of Bardeen-Cooper-Schrieffer (BCS) formalism developed earlier for the charge density wave concept. It is shown that for the combination of both the types of interaction (electron-electron and electron-phonon), the formation of a collective excitation – an electron crystal of charge density wave – in the model of exciton insulator is possible, which, in a general case, can also be accompanied by a structural transition. The results for vanadium dioxide are examined within the frameworks of the developed approach.

## 1. Introduction

The problem of metal-insulator transition (MIT) in an electric field has received increasing attention in the recent time [1-5]. This interest, on the one hand, is caused by investigations of negative differential resistance (NDR) effects in some materials exhibiting electronic switching due to the MIT. On the other hand, it is caused by investigations of the MIT phenomenon itself because the study of the electric field effect upon the parameters of the MIT would allow more detailed understanding the nature of the transition mechanism. In this work we consider the problem of the MIT occurring in an electric field, from the viewpoint of electrical switching, using the example of $VO_2$. Some of the previously obtained experimental results [6, 7] presented in figures 1- 3 can be summarized as follows. On the basis of the study of *I-V* characteristics in a wide temperature range (figure 1) it has been shown that, as the field strength increases, the transition temperature decreases (figure 2). Also, the free charge carrier density at $T_0<200$ K ($E>10^5$ V cm$^{-1}$) becomes less than the critical Mott density $n_c$ [8], which for vanadium dioxide is $n_c \sim 3 \cdot 10^{18}$ cm$^{-3}$ [5, 9] (figure 3).

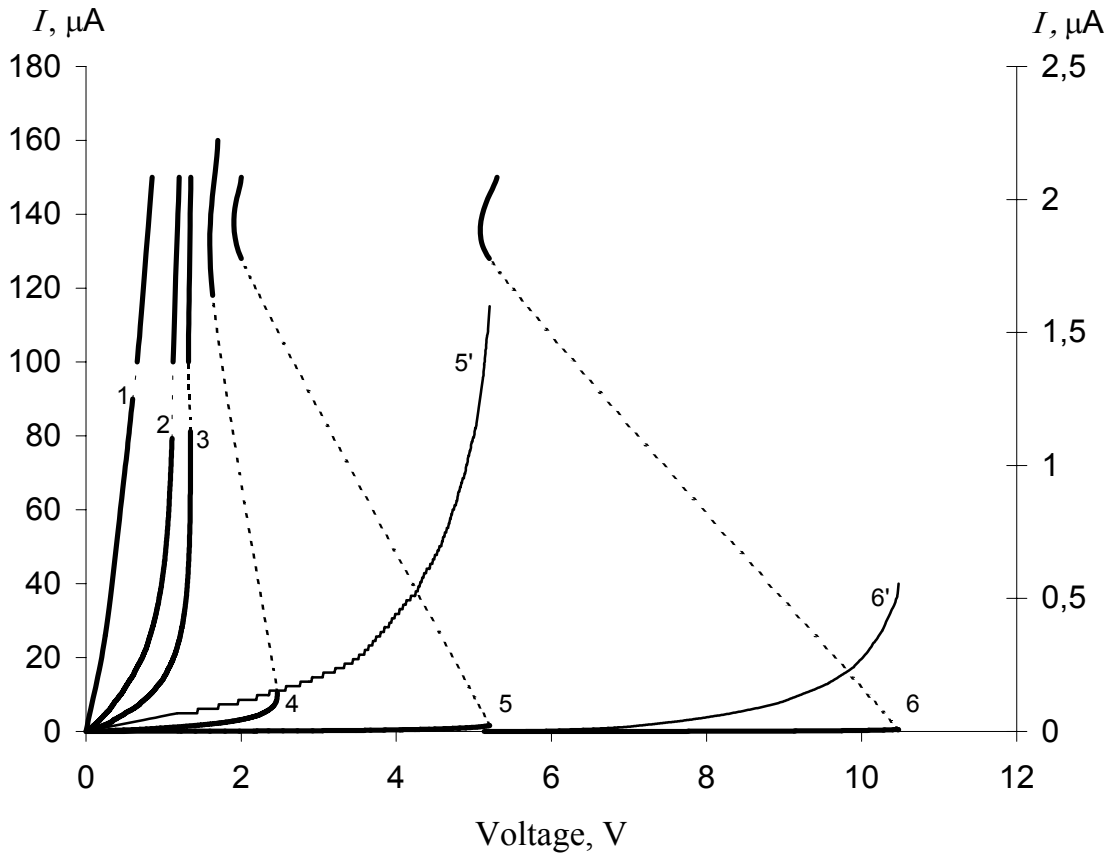

**Fig.1.** Experimental current–voltage characteristics of the VO$_2$ based switch at various ambient temperatures $T_0$ (K) [7]: (*1*) 293; (*2*) 241; (*3*) 211; (*4*) 144; (*5*) 91; (*6*) 15; (*5',6'*) low-conductivity regions of curves *5* and *6* in an extended scale (right-hand ordinate axis). Dashed lines indicate unstable (transient) regions of the *I–V* curves.

The physical mechanism of the MIT in VO$_2$ is still a matter of discussion. First of all, a fundamental question of whether this phase transition is an electronic Mott transition (and the crystal structure changes merely accompany it) is not solved yet. Also, the nature of the low-temperature (LT) insulating phase is not fully understood, i.e. it is not clear whether the LT phase is a Mott insulator or it is a conventional band insulator. These two questions are interrelated, though not equivalent. The point is that the contribution of the electron correlation effects to the energy gap might be relatively small, but these effects could nevertheless play a role of the trigger mechanism for the MIT [9]. Such a situation is not surprising, because, in general, the principle "small perturbations result in substantial changes" is the quintessence of the physics of phase transitions. The analysis of the experimental and theoretical findings (see, e.g., [8, 10-13] and references therein) leads to the following conclusions:

1) The LT phase of VO$_2$ is a band insulator, not a correlation one (that is, the energy gap is mainly determined by the crystal structure, and not by electron correlation, like it is in a typical Mott insulator, such as, e.g. NiO [8]).
2) Coulomb electron correlations play nonetheless an important role both in the insulating phase and in the metallic phase (that is, for a correct description of the electronic band structure of both the phases, the correlation effects should be necessarily taken into account).

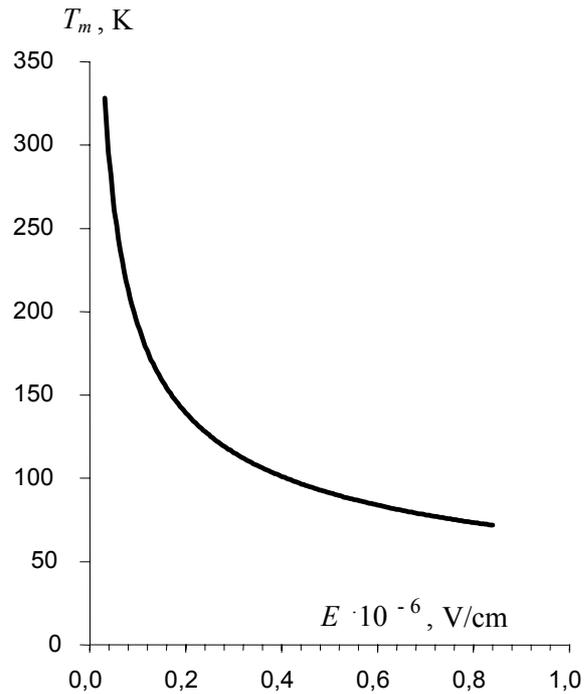

**Fig.2.** Field dependence of the critical switching temperature $T_m$ [6]. As $E \to 0$, $T_m$ tends to the equilibrium value of the MIT temperature equal to $T_t = 340$ K for vanadium dioxide [8].

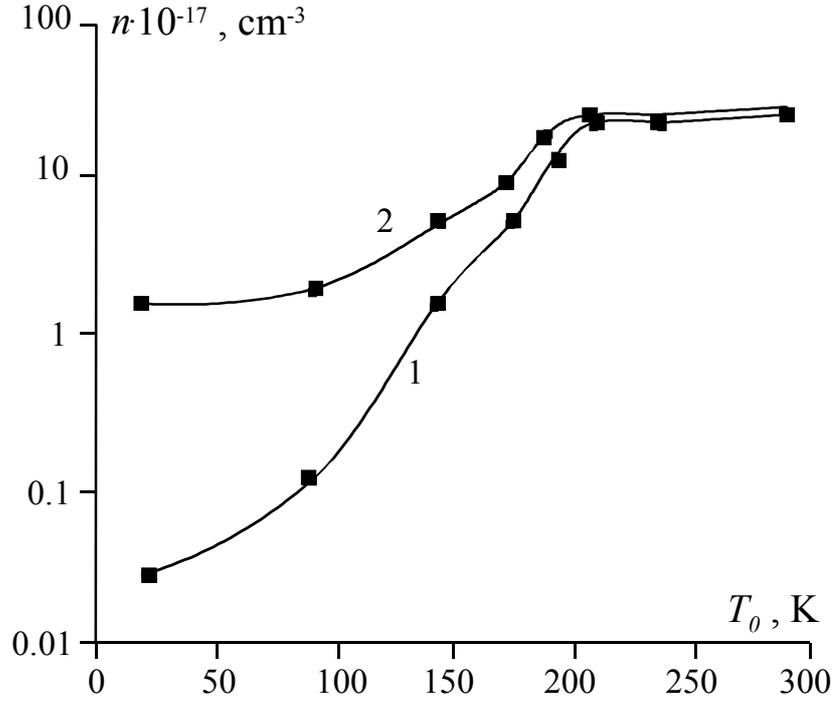

**Fig.3.** Maximum electron density in the switching cannel as a function of ambient temperature for different I-V curves in figure1; in fact, $n$ depends on the maximum electric field (threshold field $E_{th}$), and the latter, in turn, depends on $T_0$. (1) $\mu = const = 0.5$ cm$^2$ V$^{-1}$ s$^{-1}$ [8], and (2) $\mu = \mu_0 exp(-W_\mu/k_BT)$ – a typical temperature dependence for the mobility in strongly correlated systems [8, 13] with $W_\mu = 0.05$ eV for VO$_2$ (which can be estimated from the data presented in [14]). The values of $n$ were calculated from the relation $\sigma = ne\mu$, with $\sigma$ being the channel conductivity obtained from the data of figure 1 as described in [7].

In general, the question about the MIT mechanism in VO$_2$ (posed as "either Mott, or Peierls" [15]) has no sense because, in fact, this mechanism is essentially dual [9], and one can say in this case about a "Mott-Peierls transition". All the arguments *pro* and *contra* a particular "pure" mechanism are based on indirect experiments, and it is difficult – if not impossible at all – to perform an *experimentus crucis* in this situation, because it would require initiation of the MIT in a certain subsystem (electron or lattice) not touching another one. However, these subsystems are closely interrelated due to the strong electron-phonon interaction in vanadium dioxide [10].

However, we believe that the driving force of the MIT in VO$_2$ is the electron-electron interaction [9], because it has been shown [5, 9] that a necessary and sufficient condition for the transition to occur is the attainment of a certain critical concentration $n_c$. This $n_c$ may be achieved either by the equilibrium way (i.e. under heating up to $T=T_t$) or by any other way, e.g. under injection from contacts, photo-generation, or high-field ionization of the impurity levels, etc [9]. The value of $n_c$ can be found from the Mott criterion [8]:

$$a_H n_c^{1/3} \approx 0.25, \qquad (1)$$

where $a_H = \varepsilon\hbar^2/m^*e^2$ is the effective Bohr radius, $\varepsilon$ – the static dielectric permittivity (~100 for VO$_2$ [16]), and $m^*$ – the effective mass of the charge carriers (~3$m_e$ for VO$_2$ [8]).

It should be stressed that the MIT mechanism in VO$_2$ may not be considered, obviously, as a purely Mott mechanism [8-13]. The pure Mott transition occurs only at zero temperature and is controlled, e.g. in doped semiconductors, by the impurity concentration. On the other hand, the Mott-Hubbard MIT occurs in another vanadium oxide, V$_2$O$_3$, and this transition is also accompanied by a structural transformation at $T=T_t=150$ K [8, 10, 13]. The possibility of the use of equation (1) for VO$_2$ (which has been proved experimentally – see reference [5] and figure 3) just indicates the fact that the transition is electronically-induced. We also emphasize that the MIT in VO$_2$ is not, of course, a purely Peierls-like transition, because, despite of the vanadium atoms pairing along the *c* axis and the lattice period doubling in the LT phase, the material is not one-dimensional in classical sense: there is no any anisotropy of the physical properties [10].

It is pertinent to remind, however, that a structural transition accompanied by the superstructure formation might be, in principle, electronically induced too: in electron systems with screened Coulomb interaction, the insulating state with commensurate charge density waves (CDW) is analogous to an excitonic insulator (EI) (see, e.g., [10, 17], as well as the discussion in Section 3 below).

Thus, regardless of the detailed physical mechanism, the driving force of the MIT in VO$_2$ is electron-electron correlation, and hence the transition can be initiated by the introducing of additional charge carriers (electrons) at $T<T_t$. This fact is of principal importance for all our subsequent discussions concerning the MIT dynamics in vanadium dioxide under the action of electric field during switching.

## 2. Switching Mechanism Based on the Electric Field-Induced Metal-Insulator Transition

The mechanism of threshold switching has for a long time been a matter of especial concern, the principal issue being whether the mechanism is primarily thermal or electronic [18]. There have also been several works treating the switching effect as a metal-insulator transition occurring in electric field [18, 22]. The field effect upon the MIT in VO$_2$ has been studied previously, both theoretically and experimentally, in a number of works [1, 5-7, 19-22]. Particularly, a thermodynamic analysis based on the standard phenomenological approach [1, 19], using the equation for the free energy, shows that the shift of the transition temperature in electric field is

$$\Delta T_t \sim \frac{T_t \varepsilon E^2}{q}, \qquad (2)$$

where $q=250$ J cm$^{-3}$ is the latent heat of the transition [19]. The change of $T_t$ is negligible in this case (~ 1 K for $E \sim 10^5$ V cm$^{-1}$). Also, since the entropy of VO$_2$ increases at the transition into metallic phase, the value of $T_t$ increases with increasing $E$, i.e. $\Delta T_t > 0$ in equation (2) [1]. A decrease of $T_t$ in an electric field (figure 2), and finally its fall down to zero at a certain critical field $E_c$, can be obtained using a microscopic, not thermodynamic, approach based on the detailed MIT mechanism. Unfortunately, as noted in [1], we have no a quantitative theory to describe such a transition.

Nevertheless, taking into account the fact that the MIT in VO$_2$ is an electronically-induced transition, we next consider the following model. Let us imagine an array (3-dimensional lattice) of partly ionized one-electron sites with the localization radius $a_H$ and with $n$ being the number of free electrons: $0<n<n_c$. This model describes a doped semiconductor or an EI-phase at $n<n_c$, or, e.g., a material exhibiting a temperature-induced MIT, such as VO$_2$, at $T<T_t$. Application of an electric field to this system, taking also into account the current flow (since we consider the switching effect, not a stationary field effect), will result in both the thermal generation of additional carriers due to the Joule heat, and in the field-induced generation due to the autoionization caused by the Coulomb barrier lowering – a phenomenon analogous to the Poole-Frenkel effect. When the total concentration of free carriers reaches the value of $n=n_c$, the transition into the metal state (i.e. switching of the structure into the ON-state) occurs. In a material with the temperature-induced MIT, this switching (in a relatively low field) occurs just due to the heating of the switching channel up to $T=T_t$. There exists however one more possibility: the switching can occur at $n<n_c$ and $T<T_t$, but when $E=E_c$, where $E_c$ is the critical field, i.e. the field for which there are no bound states in the Coulomb potential. The upper estimate for this critical field $E_{c1}$ can be obtained from the "electron - positively charged centre" pair ionization energy $I_d$:

$$I_d = \frac{e^2}{2\varepsilon a_H}. \qquad (3)$$

Since $E_{c1} = I_d / ea_H$, therefore one can write from equation (3):

$$E_{c1} = \frac{e}{2\varepsilon a_H^2}. \qquad (4)$$

In fact, even in a field $E<E_{c1}$, the quantum autoionization (i.e. that enhanced by tunnelling) will lead to complete depletion of the electron level due to the barrier lowering in the electric field. Also, the presence of free carriers complicates the calculation of the $E_c$ value, because it requires to take the shielding effect into account, which would modify all the results obtained for the purely Coulomb potential. That is, the potential is $(-e/r)exp(-r/L_D)$, where $L_D = (\pi a_H/4k_F)^{1/2}$ is the Debye screening length (radius) and $k_F$ is the Fermi wave vector.

For switching in VO$_2$, the main mechanism for the field-induced carrier density increase is the Poole-Frenkel effect [5-7], i.e. just the field-assisted barrier lowering. In this case, the dependence of the electron concentration on temperature and field is given by the expression [23]:

$$n = N_0 \exp\left(-\frac{W - \beta\sqrt{E}}{k_B T}\right), \qquad (5)$$

where $N_0$ is a constant independent of electric field and only slightly dependent on temperature, $W$ – the conductivity activation energy, $\beta = 2\sqrt{e^3/\varepsilon}$ – the Poole-Frenkel constant, and $k_B$ is the Boltzmann constant. The maximum critical field $E_{c2}$ is that in which $W = \beta\sqrt{E}$. Thus, if $W=I_d/2$, which is the case for a uncompensated semiconductor, then

$$E_{c2} = \frac{W^2}{\beta^2} = \frac{I_d}{32ea_H} = \frac{1}{32}E_{c1}, \quad (6)$$

that is, it would be almost 30 times less than the estimate of $E_c$ from equation (4) above. In a general case, the critical field is given by the relation [24]

$$E_{c3} = \frac{\Delta}{e\xi}. \quad (7)$$

Here $\Delta$ is the bandgap width ($=2W$ for an intrinsic or uncompensated doped semiconductor), and $\xi$ is the coherence length. In terms of the EI model, these parameters are the binding energy and the size of an electron-hole pair, respectively. For $\Delta = I_d$ and $\xi = a_H$, the value of $E_{c3}$ in equation (7) coincides with $E_{c1}$ in equation (4).

Thus, there exists a certain maximum critical field $E_{c0}$, which at $T=0$ leads to a collapse of the bound exciton-like states and hence to the insulator-to-metal transition. However, in the real switching conditions ($T \neq 0$, $n \neq 0$), the actual $E_c$ is less than this $E_{c0}$ – see the discussion above, as well as the work [6], where it has been shown that the value of $E_c$ decreases with increasing temperature. Analogously, the critical field $E_c$ must decrease with increasing electron density, since, obviously, $E_c$ tends to zero at $n \to n_c$.

In the case of $n = n_c$ ($T=T_t$) and in equilibrium ($E=0$), expression (5) becomes:

$$n_c = N_0 \exp\left(-\frac{W}{k_B T_t}\right). \quad (8)$$

From expressions (5) and (8), reducing the parameter $N_o$, it is straightforward to obtain the dependence of the carrier density on field and temperature:

$$n = n_c \cdot \exp\left\{-\frac{W}{k_B T_t}\left[\frac{T_t}{T_m}\left(1 - \sqrt{\frac{E}{E_0}}\right) - 1\right]\right\}, \quad (9)$$

where $E_0=E_{c2}$ is the maximum critical field for the Poole-Frenkel effect, and $T_m=T(E)$ is the switching channel temperature which depends on $E$ (figure 2). The MIT (and switching) occurs when $n = n_c$, that is, at

$$T_m = T_t\left[1 - (E/E_0)^{1/2}\right], \quad (10)$$

when the term in square brackets in equation (9) turns into zero.

However, as was said above, apart from the condition $n = n_c$, there exists one more possibility: the MIT might occur when a field applied to the sample will reach the value of $E_c(n) < E_0$. If $E = E_0$ in equation (9), then one obtains $n = n_c \exp(W/k_B T_t) = N_0 > n_c$, but $E_0$ is the maximum critical field, and one should bear in mind that, as the field increases (and, accordingly, the value of $n$ increases too), the critical field $E_c$ changes (decreases): at $n \to n_c$, $E_c$ becomes zero, though not suddenly from its maximum value $E_0$. We surmise that this dependence might be written as

$$E_c = E_0 [1 - n/n_c]^\gamma, \qquad (11)$$

i.e. the switching event in a high field can occur at $E = E_c$ when the condition $n = n_c$ is not fulfilled yet. The unknown parameter $\gamma$ can be estimated from the scaling theory for MIT [13] and from a general view of the critical field, equation (7). Because $\Delta$ and $\xi$ can be written as [13, 25, 26]:

$$\Delta = \Delta_0 [1 - n/n_c]^\alpha, \qquad (12\text{-a})$$

$$\xi = \xi_0 [1 - n/n_c]^\nu, \qquad (12\text{-b})$$

then $\gamma = (\alpha - \nu)$. The coherence length exponent is usually in the range $\nu = -(0.5 - 2.0)$, and $\alpha > 0$ may vary within a rather wide range [25]. The data on experimental determination of the $\alpha$ and $\nu$ are not numerous in the literature. For example, in the case of n-Ge, $\alpha = 1.9$ и $\nu = -0.8$ [26]; however, even in doped semiconductors, in the vicinity of the transition point ($n \to n_c$), the $\alpha$–value may be as large as ~(5–10) – see figure 4. Therefore, a maximum value of $\gamma$ in equation (11) may reach the magnitude of about 10 or higher.

Substituting equation (11) into (9), we obtain a relation between the channel temperature and electron density at the switching event ($E = E_c$):

$$T_m = T_t \frac{1 - (1 - n/n_c)^{\gamma/2}}{1 - (k_B T_t / W) \ln(n/n_c)}. \qquad (13)$$

The graphs of these dependences are presented in figure 5, where $\gamma$ and $(k_B T_t / W)$ are considered as variable parameters.

One can see that the forms and positions of the curves in figure 5 are almost independent of the variation of the parameter $(k_B T_t / W)$ in the range 0.05 to 0.30 which corresponds to the activation energy $W \sim 0.15$–0.9 eV for $T_t = 340$ K (for vanadium dioxide, $E_g \sim 1$ eV, $W \sim 0.5$ eV, and $(k_B T_t / W) = 0.059$). As far as the $\gamma$ exponent concerned, the best accordance with the experimental data is achieved just for $\gamma \sim 10$ (cf. curves 2 and d in figure 5).

Note that equation (13), taking into account expression (11), one can re-write also as:

$$T_m = T_t \frac{1 - (E/E_c)^{1/2}}{1 - (kT_t / W) \ln[1 - (E/E_c)^{1/\gamma}]}, \qquad (14)$$

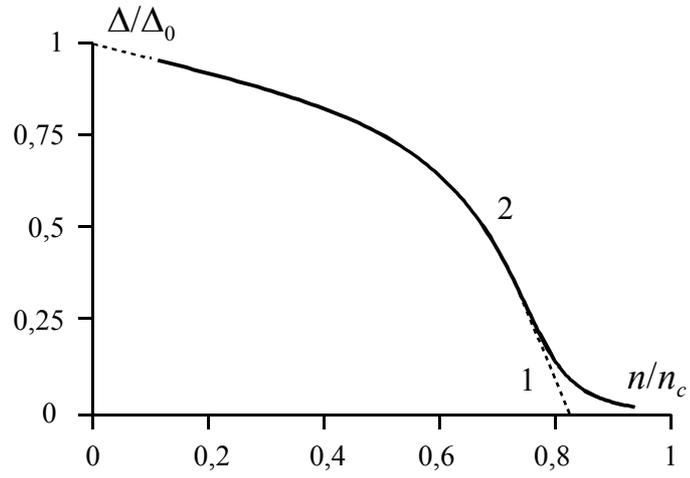

**Fig.4.** Dependence of the bandgap $\Delta$ on the electron density according to expression (12-a): (1) $\alpha < 1$, (2) $\alpha > 1$. The experimental dependence of the activation energy (which is approximately equal to $\Delta/2$) on $n$ [27] in the vicinity of the transition point corresponds to curve 2 with $\alpha \approx 5-7$. (See also the monograph [8] where these results from the work [27] has been reproduced).

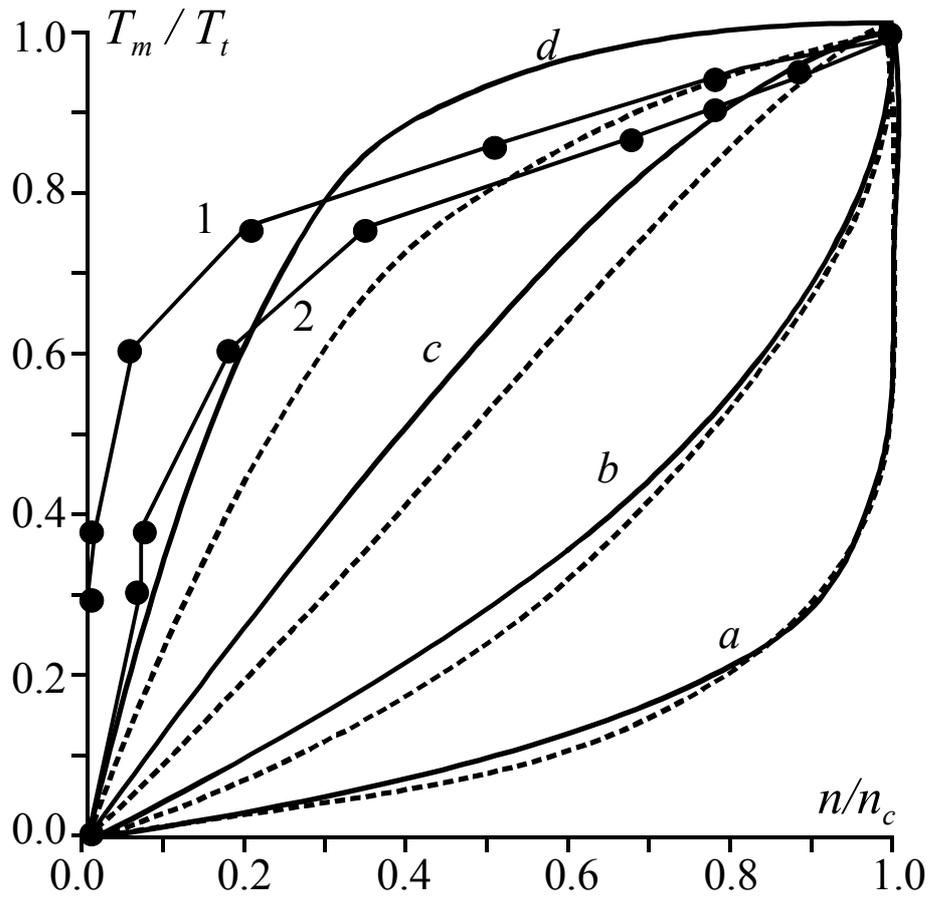

**Fig. 5.** Maximum switching channel temperature as a function of electron density at $E=E_c$ (equation (13)) with $\gamma = 0.3$ (*a*), 1 (*b*), 3 (*c*) and 10 (*d*), and $(kT_t/W) = 0.05$ (solid lines) and 0.3 (dotted lines). Experimental data (curves 1 and 2) correspond to those in figure 3.

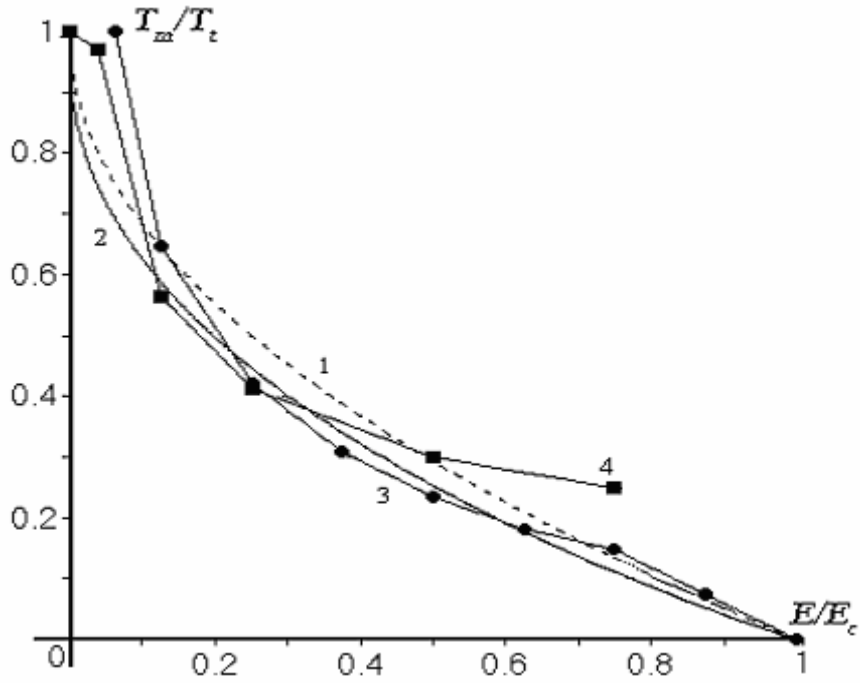

**Fig. 6.** Dependence of the transition temperature on electric field from expression (9) (1, dashed line), equation (14) with γ=10 and $(k_B T_t/W) = 0.059$ (2, solid line), and experimental data – approximate calculations [21] (3, circles) and results of numerical modelling [7] (4, squares), see figure 2.

which practically coincides with formula (10) because the denominator in equation (14) does not differ too much from unity almost in the whole range of the $E$ variation (see figure 6).

Thus, the proposed model not only allows the qualitative description of the switching mechanism, but it is in quantitative agreement with the experimental results, in particular, with those concerning the critical concentration, figures 3 and 5, and $T_m(E)$ (or $T_t(E)$) dependence, figures 2 and 6. The switching mechanism is based on the electronically-induced Mott metal-insulator transition taking into consideration the dependence of the carrier density on electric field (equation (9)), as well as the scaling of the critical field (equation (11)).

### 3. Metal–Insulator Transition with Formation of Charge Density Waves in the Model of Exciton Insulator.

As was already briefly discussed in Section 1 above, an electron-induced MIT mechanism is feasible even at the electron-phonon interaction leading to the superstructure formation (Peierls transition). This statement is based on the early work by Adler and Brooks [28] who had shown that, at a distortion of the crystal lattice in a one-dimensional system with two cations per unit cell and with one 3d electron per cation, there is the concentration dependence of the gap in the electron spectrum:

$$\Delta = \Delta_0 \left[ 1 - \beta_{e-ph} \cdot \frac{n(T)}{n_c} \right], \tag{15}$$

where $\Delta_o$ is the gap at zero temperature, and the constant $n_c$ is determined by a distortion parameter; $\beta_{e\text{-}ph}$ is a parameter proportional to the constant of electron-phonon interaction. Equation (15) differs from the above introduced scaling formula (12-a) because here we deal with a microscopic approach, and expression (15) gives solely a first (linear) member in the expansion of $\Delta$ into a Taylor's series.

Later, these ideas have been developed on the basis of the CDW concept. CDW is a specific collective excitation in the above described system. For such an excitation, the electron dispersion law must obey the condition of the electron-hole resonance (nesting) [10, 17, 29]:

$$\varepsilon_1(k+q) = -\varepsilon_2(k), \qquad (16\text{-a})$$

$$\varepsilon(k+q) = -\varepsilon(k), \qquad (16\text{-b})$$

where $q$ is a certain wave-vector and $\varepsilon_i(k)$ is the electron energy with the indexes 1 and 2 corresponding, respectively, to the conduction and valence bands; equation (16-b) is the nesting condition for the original metal phase. As such the situation corresponds either to the spherical Fermi surface or to that possessing congruent regions; this is characteristic of, e.g., $VO_2$ with $q = Q(\pi/a, 0, \pi/c)$ where $c$ is the crystal axis along which the cationic chains are directed [30].

For a narrow-band system described by the Hubbard Hamiltonian [31]

$$H = \sum_{n,\sigma}\left[(\varepsilon_1 - \mu)a^+_{n\sigma}a_{n\sigma} + \frac{U}{2}a^+_{n,\sigma}a_{n,\sigma}a^+_{n,-\sigma}a_{n,-\sigma}\right] + \\ + \sum_{n,n',\sigma} B(R_n, R_{n'})a^+_{n\sigma}a_{n'\sigma}, \qquad (17)$$

with the spectrum (16-b), taking into account the electron-phonon interaction

$$H_{e-ph} = \sum_{k,q,\sigma}\eta(q)(b(q)+b^+(-q))a^+_\sigma(k)a_\sigma(k-q), \qquad (18)$$

the self-consistent equation for the gap (analogous to the BCS equation in the theory of superconductivity) has been obtained [32]:

$$\lambda^{-1} = \int_0^B d\varepsilon \frac{1}{\sqrt{\varepsilon^2+\Delta^2}} \tanh\left[\frac{\sqrt{\varepsilon^2+\Delta^2}}{2k_B T}\right]. \qquad (19)$$

Here $\varepsilon$ is the energy of a single-electron atomic level, $\mu$ – the chemical potential, $U$ – on-site intra-atomic repulsion energy, and $a^+_{n,\sigma}(a_{n,\sigma})$ are the creation (annihilation) operators on a site in the Wannier representation with the translation vector $R_n$, and $B(R_n, R_{n'})$ is the overlap integral of wave functions (the transfer integral), $B$ – the band width.

$$a^+_\sigma(k) = \frac{1}{\sqrt{N}}\sum_n a^+_{n,\sigma}\cdot \exp(ikR_n) \text{ and } a_\sigma(k) = \frac{1}{\sqrt{N}}\sum_n a_{n,\sigma}\cdot \exp(ikR_n) \qquad (20)$$

are the creation and annihilation operators in the Bloch representation, $\eta(q)$ and $b^+(q)$ ($b(q)$) are the electron-phonon interaction constant and the phonon creation (annihilation) operators, respectively, $N$ – the number of atoms, $\sigma$ – the spin projection (up or down). The quantity $\lambda = (4\eta(Q)/\omega_Q - U) N_\rho(0)$ ($\omega_Q$ – the phonon frequency, $N_\rho(0)$ – the Fermi level density of states) determines a singlet (CDW) solution. The dependence of the gap $\Delta$ on the carrier density in the model of EI has been earlier obtained in the work [33].

The temperature dependence of the gap in equation (19) is, in fact, the dependence $\Delta(n(T))$. As the temperature increases, the thermally-excited electrons and holes appear, and this increase in the carrier density leads to a reduction of the CDW amplitude. As a result, the gap (which is proportional to this amplitude) decreases leading to further increase of the number of excited electrons, further diminishing of the gap, etc. Then, at $T = T_t$ the gap eliminates and the second-order insulator-metal transition occurs accompanied by the CDW disappearance.

A sharp transition into the metallic state, corresponding to a first-order phase transition, i.e. to switching in $VO_2$, is possible at a violation of the resonance condition (16), when some other bands, along with the congruent regions of the $\varepsilon(k)$, are present; these bands do not possess any peculiarities and do not take part in the CDW formation. For example, in the Hubbard model, one can take into consideration the overlap of the electron functions (electron hoping) between the neighbour vanadium atoms next after the nearest ones. The latter condition leads to the Fermi surface modulation, i.e., in fact, to the CDW amplitude change. In this case, the CDW deformation in, e.g., an electric field, might result in the change of the free electron density, and the self-consistency equation will thus describe the dependence of the gap width on the field [6] (see figure 7), not only the dependence on the temperature. On the other hand, this self-consistent mechanism of the gap narrowing and the free carrier thermal activation is akin, to some extent, to the Poole or Poole-Frenkel effects with some regions of the electron spectrum playing the role of "impurity centres". These regions do not possess the property (16) and merely play the role of a reservoir (in $k$-space) for electrons and holes.

For the state of EI, the singlet dielectric coupling leading to the CDW state forms also in the case of an electron-electron interaction with a Hamiltonian [29, 32]:

$$H_{e-e} = \sum_{k,k',q,\sigma} g(q) a^+_{1,\sigma}(k) a^+_{2,\sigma}(k') a_{2,\sigma}(k'+q) a_{1,\sigma}(k-q) , \qquad (21)$$

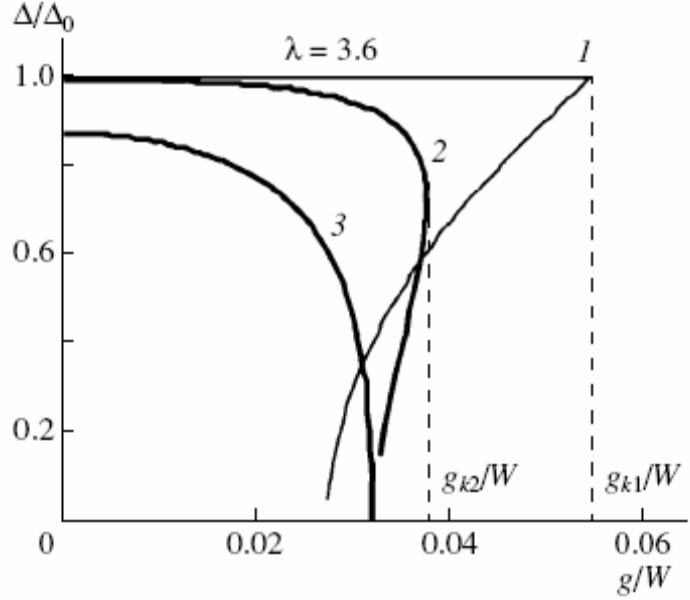

**Fig.7.** Bandgap width Δ as a function of the parameter g plotted in the normalized coordinates for different values of $k_B T/W$: (*1*) 0.0001, (*2*) 0.01, and (*3*) 0.02 [6]. The parameter g determines the Fermi surface modulation and depends linearly on the electric field *E*; $g_k$ is a critical value of g corresponding to the critical field $E_c$, and λ is the parameter of equation (19).

where *g(q)* is constant of the screened 'density-density' Coulomb interaction, $a^+_{i,\sigma}(k)(a_{i,\sigma}(k))$ are the creation (annihilation) Fermi operators for the conduction (1) and valence (2) bands respectively, and the spin projection σ determines the interaction without the spin overturn. Like in the case of the electron-phonon mechanism, the nesting condition, equation (16-a), determines here a preferential scattering with *q = Q* because the scatter amplitude of an electron with the wave-vector *k* on a hole with the wave-vector (*k + Q*) has a logarithmic singularity with an imaginary pole. This singularity characterizes instability of the system with respect to the electron-hole pairing even at a very weak (strongly screened) Coulomb attraction ($na_H^3 \gg 1$) [29].

The 'density – density' interaction (21) forms an electron spectrum described by an equation similar to equation (19) [32]. The CDW state is due to the correlation interaction between the bands. This interaction is the Coulomb attraction between electrons and holes, and as the temperature decreases, the electron-hole coupling becomes thermodynamically advantageous (the Mott exciton formation). At this point, a second-order phase transition occurs.

Using the transformation described by equation (20), it is straightforward to show that both the terms (18) and (21) yield identical contributions to the Hubbard model corresponding to the pairing of electrons from the two bands with the wave-vector *q=Q*, due to either electron-phonon or electron-electron interaction:

$$H_{e-ph} = \sum_{n,\sigma} \eta(\mathbf{Q})(b_Q + b^+_{-Q}) a^+_{n,\sigma} a_{n,\sigma} \exp(-i\mathbf{Q}\cdot\mathbf{R}_n), \qquad (22)$$

$$H_{e-e} = \sum_{n,n',\sigma} g(Q) a^+_{1n,\sigma} a^+_{2n',\sigma} a_{2n',\sigma} a_{1n,\sigma} \cdot \exp(-iQ(R_n - R_{n'})). \qquad (23)$$

The contribution (22) forms a spectrum with CDW, using uncoupling on mean values of phonon and electron operators [32]:

$$< b_Q a^+_{n,\sigma} a_{n,\sigma} > \approx < b_Q > \cdot < a^+_{n,\sigma} a_{n,\sigma} > \qquad (24)$$

The similar result turns out to be for the contribution (23), when one uses the uncoupling on mean values of electron operators from different bands [29]:

$$< a^+_{1n,\sigma} a^+_{2n',\sigma} a_{2n',\sigma} a_{1n,\sigma} > \approx n_{1\sigma} \cdot n_{2\sigma} \qquad (25)$$

where $n_{i\sigma} = < a^+_{i\sigma} a_{i\sigma} >$.

Thus, for the combination of both the types of interaction, the formation of a collective excitation – an electron crystal of CDW – in the EI model is possible, which, in general, can also be accompanied by a structural transition. For example, in the case of vanadium dioxide, there is a transformation from the LT monoclinic structure to the high-T tetragonal one.

Note that, in the presence of both the interaction types, their individual contributions to the formation of the dielectric gap in the spectrum are not independent. If the gap $\Delta$ appears due to interaction (21), then the anomalous average $< a^+_{1,\sigma}(k) a_{2,\sigma}(k+Q) >$ (singlet coupling) causes the lattice deformation because of the electron-phonon interaction. The deformation is proportional to the term $< b_Q + b^+_{-Q} >$, which, in turn, is proportional to $\Delta$. Thus, the loss in the deformation elastic energy, associated with the appearance of non-zero average $< b_Q + b^+_{-Q} >$, is compensated by the gain in the electron energy due to the simultaneous appearance of the gap because of the singlet coupling.

## 4. Conclusion.

In conclusion, we emphasize once again the main idea of the proposed switching mechanism. This mechanism is based on an electronically-induced transition (i.e., in fact, the Mott-Hubbard MIT) occurring in conditions of the non-equilibrium carrier density excess in the applied electric field. In Section 2, this mechanism was developed on the basis of a rather phenomenological approach. This model not only allows the qualitative description of the switching mechanism, but it is in quantitative agreement with the experimental results, in particular, with those concerning the critical concentration and the $T_t(E)$ dependence. The model takes into account the dependence of the carrier density on electric field, and scaling of the critical field.

In Section 3, we have attempted to show that such a "macroscopic" (semi-phenomenological) approach can be supported by some microscopic model. The quintessence of this approach consists in the fact that an electronically induced MIT might, in principle, be described in terms of BCS formalism developed earlier for the CDW concept. It was particularly shown that for the combination of both the types of interaction (electron-electron and electron-phonon), the formation of a collective excitation – an electron crystal of CDW – in the model of exciton insulator is possible, which, in general, can also be accompanied by a structural transition.

In the present work, these models have been developed for a particular case ($VO_2$), but they are also applicable to other materials (such as, e.g., $NbO_2$ which exhibits the MIT at $T_t$ = 1070 K [8, 13, 34, 35]), and even to the materials exhibiting only the non-equilibrium MIT, for example, to amorphous semiconductors (including chalcogenide glass semiconductors, CGS) [18, 35, 36]. Moreover, the model can describe the switching effect with N-type NDR, not only that with S-type NDR. For the former, switching is associated with the inverse MIT, when the low-temperature phase is metallic. Such re-entrant MITs have been observed in $V_2O_3$:Cr, $NiS_2$:Se, EuO, CMR-manganites, and some other materials [8, 13, 37]. N-type switching has been experimentally studied in the structures based on $V_2O_3$:Cr [38], $NiS_2$:Se [39], $Sm_{1-x}Sr_xMnO_3$ [40]. A possible mechanism of the electronically-induced inverse MIT has been discussed in [37], where the temperature-induced inverse Mott transition in EuO and CMR-manganites, depending on the dopant concentration, is considered. The switching mechanism in this case may be associated with the carrier extraction from the metal phase similarly to the switching effect in materials with the standard (not inverse) MIT due to the carrier injection into the semiconducting phase as was discussed above.

Thus, the main features of switching both in transition metal compounds and in other strongly correlated systems can be naturally explained within the frameworks of a universal switching mechanism based on the electronically-induced metal-insulator transition. In some cases, a preliminary electroforming is required. The switching channel forms in the initial structure during such electroforming resulted from the electro-thermal and electrochemical processes under the action of the applied forming voltage [18, 22, 36]. This channel consists (partly or completely) of a material [22] which can undergo a MIT from one stable state into another at a certain critical temperature $T_t$ or electron density $n_c$. The S- or N-shaped *I-V* curve is conditioned by the development of an electro-thermal instability in the switching channel. Due to the effect of Joule heating, when the voltage reaches a critical value $V=V_{th}$, the channel is heated up to $T=T_t$ and the structure undergoes a transition from the insulating OFF to the metallic ON state (for the case of S-switching). This is the model of "critical temperature" (i.e. a simple electrothermal mechanism, albeit taking into account the specific $\sigma(T)$ dependence of the material at the MIT), though the mechanism of the MIT itself is, of course, essentially electronic. When the ambient temperature $T_0$ is much less than the transition temperature, and the value of $E_{th}$ is high enough, the effect of electronic correlations upon the MIT is feasible. In high electric fields electronic effects influence the MIT so that a field-induced increase in the charge carrier concentration (due to either injection from contacts or impact ionization, or due to field-stimulated donor ionization – i.e. the Poole-Frenkel effect) leads to the elimination of the Mott-Hubbard energy gap at $T < T_t$ [22, 34]. This effect may be treated also as a lowering of $T_t$ due to an excess negative charge (electrons), and the dependence of $V_{th}$ on $T$ deviates from the behaviour described by the critical temperature model. In this case, switching commences when a certain critical electron density $n_c$ is achieved. In the equilibrium conditions, the value of $n_c$ is achieved merely due to the thermal generation of carriers at Joule heating up to $T \sim T_t$. Also, in the even higher fields, switching occurs at $E=E_c$, when $T_m<T_t$ and $n<n_c$.

The transition of this type is important in those materials where the usual temperature-induced MIT with a well defined $T_t$ does not take place, for example, in CGS-based switching structures (for which, obviously, the equilibrium transition temperature $T_t > T_g$, while $T_t(E) < T_g$, where $T_g$ is the temperature of the glass-crystal transition) [36]. Thus, the proposed model of the switching mechanism, being a combination of the electro-thermal model of "critical temperature" and the theory of "electronic phase transition" [18], can also be valid for the description of the switching phenomena in various materials, including, e.g., CGS. In some cases, the MIT may be unobservable under the temperature change in equilibrium conditions, but it displays itself only under the applied electric field. It is quite likely that such a situation just takes place in CGS; similar effects obviously occur also in some MOM structures based on transition metal oxides [22, 34-36, 38-40].